\documentclass[ag]{copernicus}

\usepackage{color}

\begin{document}

\title{{QM theory of the thermal electron gyroradius}}

\author[1,2]{{R. A. Treumann}
}
\author[3]{{W. Baumjohann}}

\affil[1]{Department of Geophysics and Environmental Sciences, Munich University, Munich, Germany}
\affil[2]{International Space Science Institute, Bern, Switzerland}
\affil[3]{Space Research Institute, Austrian Academy of Sciences, Graz, Austria}

\runningtitle{Gyroradius}

\runningauthor{R. A. Treumann and W. Baumjohann}

\correspondence{R. A. Treumann\\ (rudolf.treumann@geophysik.uni-muenchen.de)}

\received{ }
\revised{ }
\accepted{ }
\published{ }


\firstpage{1}

\maketitle

\begin{abstract}
The average (thermal) gyroradius of charged particles is re-examined from a quantum-mechanical point of view. The straight quantum-mechanical calculation clearly reproduces its conventionally used random-mean-square (rms) expectation value. The quasi-classical approach reproduces its mean expectation value as well thus confirming the purely classical calculation. It shows that the fluctuations in the gyroradius amount to 21.5\% of its rms value. This fluctuation is, however,  within the usual $\sqrt{2}$-``freedom'' range of choice in the definition of the rms gyroradius respectively the mean thermal speed on which the rms value is based, its correct (nonrelativistic) value $v_t^2= 2T_\perp/m_e$ and its simplified version $v_t^2=T_\perp/m_e$.

\end{abstract}

\introduction
The gyroradius $\rho$ of a particle of charge $e$ and mass $m$ in a magnetic field of strength $B$ is one of the fundamental parameters used in plasma physics. Its exact classical definition for a single particle of perpendicular velocity $v_\perp$ is $\rho_e = v_\perp/\omega_{ce}$ written here for an electron with cyclotron frequency $\omega_{ce}=eB/m_e$ \citep[see any textbook on plasma physics, e.g.,][]{treumann1996}. Its ion equivalent is obvious. Single electrons have undefined temperature, usually assumed to be zero, hence the exactness of the given expression. In quantum mechanics electrons obey Landau levels possessing a well defined magnetic length $\lambda_m=\sqrt{2\pi\hbar/eB}$ which corresponds to the gyroradius of an electron in the lowest Landau energy level, $\epsilon_{LLL}=\frac{1}{2}\hbar\omega_{ce}$. 

In high temperature plasmas one defines the gyroradius through the thermal velocity, respectively the perpendicular temperature $T_{\perp}=\frac{1}{2}m_e v_{t\perp}^2$. A more precise definition makes use of the velocity distribution function of the electrons to calculate the average gyroradius  $\langle\rho_e\rangle$ as the moment of the classical distribution with respect to the perpendicular particle velocity. In the random-mean-square (rms) limit this yields the above expression for $\rho_{e,\,\mathrm{rms}}$ with replacement $v_\perp\to v_{t\perp}=\sqrt{2T_\perp/m_e}$, which is nothing but the well-known rms ``thermal gyroradius". Occasionally freedom is taken for convenience in either keeping or dropping the $\sqrt{2}$ value in the definition of the thermal speed which introduces a 29\% indeterminacy of the thermal speed and rms gyroradius. 

For temperatures $T_\perp\sim 0$, i.e. $T_\perp\lesssim \epsilon_F$, with $\epsilon_F$ Fermi temperature,  electrons become degenerate and one needs to refer to the Fermi distribution. In this case the thermal speed is replaced by the Fermi velocity $v_F=\sqrt{2\epsilon_F/m_e}=(\hbar/m_e)(6\pi^2 N)^{1/3}$, a \emph{collective} velocity of degenerate electrons. One may note that this does  \emph{not} reproduce the above quantum mechanical magnetic length of electrons, $\lambda_m$!  The latter being a single electron property in magnetic fields caused by the quantum character of the magnetic flux with flux quantum $\Phi_0=2\pi\hbar/e$, not a collective effect. It defines the radius of a magnetic field line through $B=\Phi_0/\pi\lambda_m^2$. At non-zero temperatures $T_\perp\gtrsim \epsilon_F$ slightly exceeding $\epsilon_F$ a small correction is to be introduced on the Fermi velocity entering the collective rms gyroradius.

In many particle theory, starting from the single particle picture one should, for greater precision, make use however of the distribution of particles over the entire spectrum of energy states available and accessible to the population at a given temperature $T_\perp$. 

Since only two particles of opposite spins can occupy any energy state there is a well-known competition between the available states and the number of particles at $T_\perp$. If the temperature $T_\perp\lesssim T_F$, with $T_F$ the Fermi temperature, then the particles have effectively zero temperature, and the gyroradius is defined through their single velocity in the respective state.  This case is encountered in high density compounds at
\begin{equation}\label{denslim}
N_e\sim \left(\frac{m_eT_\perp}{2\pi^2\hbar^2}\right)^\frac{3}{2}\approx  4.6\times10^{11} T_{\perp,\mathrm{[eV]}}^\frac{3}{2} \quad \mathrm{m}^{-3}
\end{equation}
with temperature measured in eV. (One may note that this number is remarkably close to the conditions encountered in the solar chromosphere!). When this condition holds, quantum effects become important; it implies that the interparticle distance
\begin{equation}
N_e^{-\frac{1}{3}}\simeq \lambda_T, \qquad \lambda_T=\left(2\pi\hbar^2/mT_\perp\right)^\frac{3}{2}
\end{equation}
equals the thermal de Broglie wavelength $\lambda_T$, here written with $T=T_\perp$.
 At higher temperatures and lower densities the average gyroradius should be calculated by adding up all electrons in the available states. By assuming a classical distribution function this is circumvented and reduced to a simple integration yielding the above thermal rms value $\rho_{e,\,\mathrm{rms}}=\sqrt{2m_eT_\perp}/\omega_{ce}$. Here we present the exact quantum mechanical calculation which naturally confirms the thermal rms gyroradius result but, in addition, yields the expectation value $\langle\rho_e\rangle$ of the gyroradius and provides its width of uncertainty, i.e its thermal spread. 

\section{Quantum mechanical treatment}
Knowing the spectrum of quantum mechanical states of an electron in magnetic field there are two ways of calculating the thermal gyroradius at given temperature $T_\perp$. Either the gyroradius is calculated by averaging it over the distribution -- which is its expectation value $\langle\rho_e\rangle$, would be the exact way to do it --, or one calculates the energy expectation value $\langle\epsilon_\perp\rangle$ which provides the expectation value of the squared gyroradius $\langle\rho_e^2\rangle$ whose root can also be taken to represent the thermal gyroradius. This latter value will be calculated first.

\subsection{Root-mean squared gyroradius}
The energy levels of an electron in a homogeneous magnetic field (the case of interest here as a magnetic inhomogeneity provides only higher order corrections) has been calculated long ago \citep{landau30,landau1965}. Since the parallel dynamics of an electron is not of interest here, it can be taken classically. Then the energy of an electron in the magnetic field becomes
\begin{equation}
\epsilon_e=\epsilon_\| +\epsilon_\perp(\ell,\sigma)=\epsilon_\|+\hbar\omega_{ce}\left(\ell+{\textstyle\frac{1}{2}+\sigma}\right), \qquad \ell\in\mathsf{N}
\end{equation}
with $\epsilon_\|=p_\|^2/2m_e$, quantum number $\ell$, and $\sigma=\pm\frac{1}{2}$ the two spin directions. The average distribution of electrons over these $\ell$ perpendicular energy states is given by the Fermi distribution
\begin{equation}
\langle n_F\rangle =\frac{1}{1+z^{-1}\exp(\beta_\|\epsilon_\|+\beta_\perp\epsilon_\perp)}, \quad z=\mathrm{e}^{\beta_\perp\mu_\perp+\beta_\|\mu_\|}
\end{equation}
written here for the anisotropic case, with $z$ the fugacity which depends on the chemical potentials $\mu_{\|,\perp}$ in parallel and perpendicular direction, and $\beta_{\|,\perp}=T_{\|,\perp}^{-1}$ are the inverse temperatures (in energy units).\footnote{One may notice that this is the \emph{only} compatible with statistical mechanics anisotropic form of the Fermi distribution. It shows that both, the chemical potential and inverse temperature are vector quantities, i.e. in relativistic terms they are 4-vectors, a conclusion which has been much debate about in relativistic thermodynamics \citep[for references see][]{nakamura2006} but here comes out naturally without much effort.} At sufficiently large temperatures the unity in the denominator is neglected, a very well justified approximation which still takes into account the non-continuous energy distribution over discrete Landau levels thus maintaining the quantum character of electrons. The fugacities enter into the normalization constant now. This is the case far away from Eq. (\ref{denslim}) which for plasmas  interests us here. Under these conditions the expectation value of the (average) perpendicular energy of the electrons (i.e.the perpendicular electron pressure)  is calculated from the integral
\begin{equation}
\langle N\epsilon_\perp\rangle\propto \int\limits_{-\infty}^\infty dp_\|\mathrm{e}^{-\beta_\|\epsilon_\|}\!\!\sum\limits_{\sigma=\pm \atop\ell=0}^\infty \epsilon_\perp(\ell,\sigma)\ \mathrm{e}^{-\beta_\perp\epsilon_\perp(\ell,\sigma)}
\end{equation}
The spin contribution in the perpendicular energy either compensates for the half Landau energy level or completes it to the first order level. Thus the sum splits into two terms which both are geometric progressions which can immediately be done. The final result, using the normalization of the integral to the average density of particles and dividing through $N=\langle N\rangle$ thus yields for the average energy
\begin{equation}
\langle\epsilon_\perp\rangle = \frac{2\beta_\perp(\hbar\omega_{ce})^2\mathrm{e}^{-\beta_\perp\hbar\omega_{ce}}}{\left(1-\mathrm{e}^{-\beta_\perp\hbar\omega_{ce}}\right)^2}\Big(1-{\textstyle\frac{1}{2}}\mathrm{e}^{-\beta_\perp\hbar\omega_{ce}}\Big)
\end{equation}
At the assumed large temperatures the exponentials must be expanded to first order yielding the very well known and expected classical result that the average energy is the temperature, $\langle\epsilon_\perp\rangle=\beta^{-1}_\perp\equiv T_\perp$. Hence, taking its root and inserting into the gyroradius we find what is expected in this case:
\begin{equation}
\rho_{e,\mathrm{rms}}=\sqrt{\langle\rho_e^2\rangle}={\sqrt{2m_e\langle\epsilon_\perp\rangle}}/{eB}, \qquad \langle\epsilon_\perp\rangle=T_\perp
\end{equation}
This is the root-mean-square gyroradius, a well known quantity. At lower temperatures $T_\perp\gg \epsilon_F$, still by far exceeding Fermi energy. the former expression for $\langle\epsilon_\perp\rangle$ has to be used in this expression.

\subsection{Expectation value}
However, the correct gyroradius is not its root mean square but the expectation value $\langle\rho\rangle$. This is substantially more difficult to calculate. There are two ways of finding the expectation value. Either one makes use of the Landau solution of the electron wave function and refers to the Wigner distribution of quantum mechanical states. This, for our purposes would be overdoing the problem substantially. We do not need the quantum mechanical probability distribution for the simple estimate we envisage here. Afterwards we would anyway have to combine the Wigner function with a suitable momentum distribution. The second and simpler way is to refer to the above Fermi distribution and directly using the energy distribution as above.  Under the same conditions as in the calculation of the rms value, this procedure circumvents the use of  the wave function being interested only in the energy levels. It, however, requires the solution of the integral-sum
\begin{equation}
\langle N\rho_e\rangle\propto\int\limits_{-\infty}^\infty dp_\|\mathrm{e}^{-\beta_\|\epsilon_\|}\sum\limits_{\sigma=\pm,\ell=0}^\infty\frac{\sqrt{\epsilon_\perp(\ell,\sigma)}}{\omega_{ce}}\mathrm{e}^{-\beta_\perp\epsilon_\perp(\ell,\sigma)}
\end{equation}
The sum, the integral contains, cannot anymore be done in closed form as there is no known way to tackle the summation over the root quantum index $\ell$ in a non-geometric progression. (We may note in passing that for a non-moving plasma calculating the average velocity moment would lead to a null result. However, in calculating the expectation value of the gyroradius, the perpendicular velocity, being a radius vector in perpendicular velocity space, is always positive. This is reflected by the summation over positive quantum numbers $\ell$ only. It and the gyroradius are positive definite quantities which both do not average out.) One thus needs to deviate from summing and to approximate the sum by an integral, which slightly overestimates the final result. Transforming the energy into a continuous variable, which implies that the summation index becomes continuous, then simply leaves us with a Gaussian integral of which we know that the mean value and the rms values are related by the classical formula
\begin{equation}
\langle\rho_e\rangle \lesssim 0.886\sqrt{\langle\rho_e^2\rangle}=0.886\,\rho_{e,\mathrm{rms}}
\end{equation}
the classical result for the mean, where by the $\lesssim$ sign we indicated  that the integral yields an upper limit for the expectation value of the gyroradius.

\subsection{Fluctuation}
The above estimates permit determining the thermal fluctuation (thermal spread) of the gyroradius. This fluctuation is the difference between the mean square and squared expectation values. With the above results we find that the thermal fluctuation of the gyroradius amounts to
\begin{equation}
\Delta\langle\rho_e\rangle^2\approx \langle\rho_e^2\rangle -\langle\rho_e\rangle^2\approx 21.5\%\langle\rho_e^2\rangle
\end{equation}
This number falls into the $\sqrt{2}$ range of ``freedom'' in choosing the thermal speed that is commonly used in the definition of the rms gyroradius as either $v_t=\sqrt{2T/m_e}$ or $v_t=\sqrt{T/m_e}$. 

\section{Discussion}
The expected result of the quantum mechanical calculation of the {electron} gyroradius of a hot plasma is that the root-mean-square value of the gyroradius is reproduced by quantum mechanics in the high temperature limit which clearly completely justifies its use not adding anything new except for the confirmation of well-known facts which can be generated in much simpler way by using the Boltzmann calculation. For lower temperatures we obtained a slightly more precise expression for the average energy which should be used in the definition of the thermal gyroradius. The expectation value of the gyroradius is different from the root mean square value, which is also known. Its value is lower given by the Gaussian mean. Usually this value is not used in any classical calculation. The difference between both values is within the commonly used $\sqrt{2}$ freedom for choosing the rms thermal speed.

The above calculation is based on keeping the discrete distribution of all plasma electrons over Landau levels. Since these are equally spaced up to infinity, the energy remains discontinuous even at large temperatures. The correction is, however, unimportant. 

It is sometimes claimed that the discrete distribution of electron energies is reflected in the dielectric function of a plasma which leads to collective cyclotron harmonics, i.e. Bernstein modes. However, Landau levels have nothing in common with cyclotron harmonics. They are single particle energy levels occupied at most by two electrons of oppositely directed spins, while cyclotron harmonics are eigenmodes of the magnetized plasma, channels which allow for propagation of electro(magnetic) signals. The discrete energy distribution of the electrons naturally disappears in the cacluation of the expectation values and is not reflected in any cyclotron harmonic waves. The average gyroradius appearing in the arguments of the modified Bessel functions in the dielectric expression of the harmonics in the combination $k_\perp\rho$ is a \emph{mere scaling factor} of wavelengths related to the perpendicular thermal speed. It cannot be taken as definition of a gyroradius, which is a particle property. In fact, calculation of the dielectric function from quantum mechanics accounting for the Landau energy distribution is very difficult. In classical theory one integrates the Vlasov equation with respect to time along classical particle orbits $\mathbf{v}(x,t),\, \mathbf{x}(t)$ which, in quantum mechanics, is forbidden as \emph{no orbits exist}. Orbits are replaced by probability distributions in space varying with time. The calculation therefore needs to be performed via a Feynman path integral approach which has not yet been done. It is, however, reasonably expected that it will, in the classical limit reproduce the classical dielectric function and the harmonic structure of Bernstein and cyclotron modes.

\begin{acknowledgements}
This research was part of an occasional Visiting Scientist Programme in 2006/2007 at ISSI, Bern. RT thankfully recognizes the assistance of the ISSI librarians, Andrea Fischer and Irmela Schweizer.
\end{acknowledgements}

\end{document}